\documentstyle [11pt]{article}
\begin{document}
\newcommand{\JPA}[3]{J.~Phys.~A {\bf #1}, #2 (#3)}
\newcommand{\PRL}[3]{Phys.\ Rev.\ Lett.\ {\bf #1}, #2 (#3)}
\newcommand{\PRB}[3]{Phys.\ Rev.\ B {\bf #1}, #2 (#3)}
\newcommand{\al}{\alpha}\newcommand{\bet}{\beta}
\newcommand{\be}{\begin{equation}}\newcommand{\ee}{\end{equation}}
\title{Scaling and infrared divergences in the replica field
theory of the Ising spin glass}
\author{T.~Temesv\'ari\\Institute for Theoretical Physics,
E\"otv\"os University,\\H-1117 P\'azm\'any P\'eter s\'et\'any 1, Bld.~A,
Budapest, Hungary
\and
        C.~De Dominicis\\Service de Physique Th\'eorique,
	CEA Saclay,\\F-91191 Gif sur Yvette, France
\and    I.~Kondor\\Department of Physics of Complex Systems,
E\"otv\"os University,\\
 H-1117 P\'azm\'any P\'eter s\'et\'any 1, Bld.~A,
Budapest, Hungary}
\maketitle
\begin{abstract}
Replica field theory for the Ising spin glass in zero magnetic field
is studied around the upper critical dimension $d=6$. A scaling theory
of the spin glass phase, based on Parisi's ultrametrically organised
order parameter, is proposed. We argue that this infinite step replica
symmetry broken (RSB) phase is nonperturbative in the sense that
amplitudes of scaling forms cannot be expanded in term of the coupling
constant $w^2$. Infrared divergent integrals inevitably appear when we
try to compute amplitudes perturbatively, nevertheless the $\epsilon
$-expansion of critical exponents seems to be well-behaved. The origin
of these problems can be traced back to the unusual behaviour of the
free propagator having two mass scales, the smaller one
being proportional to the perturbation parameter $w^2$ and providing
a natural infrared cutoff. Keeping the free propagator unexpanded
makes it possible to avoid producing infrared divergent integrals.
The role of Ward-identities and the problem of the lower critical
dimension are also discussed.
\end{abstract}

Spin glasses \cite{Review}
have posed a formidable theoretical challenge
ever since Edwards
and Anderson \cite{EA} proposed their simple looking model containing
the two basic features of disordered  systems: randomness and frustration.
Sherrington and Kirkpatrick (SK)
\cite{SK} defined
the mean field version of
this model on a fully-connected lattice where each
spin interacts with all the others. The highly non-trivial
solution of this mean field model by Parisi 
\cite{Pa1} is now generally accepted as the correct one. 

There is, however, no consensus on whether
the picture of ultrametrically organised pure states emerging
from Parisi's solution survives in finite dimensional,
short range systems.
"Droplet theory" \cite{McMillan,droplet1,droplet2} 
claims the opposite: in $d$
dimensions the equilibrium spin glass phase is unique, apart
from reflections of the Ising spins, and a transition from the
paramagnet to the spin glass takes place only in zero magnetic
field. A huge amount of numerical work has been devoted to clarifying
this point (see \cite{num} and references therein) and, in contrast to
droplet theory, a complex phase space structure and 
ultrametricity seem to emerge in four dimensions.
The interpretation
of recent rigorous results of mathematical physics 
\cite{NeSt,Guerra} is ambiguous, and what we can learn from
them is that even the definition of some quantities like
the probability distribution of overlaps, $P(q)$,
is a difficult problem, and that chaotic size dependence can render
the thermodynamic limit meaningless in some cases. We feel that
our findings below
about the unavoidable infrared problems of replica
field theory, arising  from a blind application of
perturbation expansion,
are somehow related to these phenomena.

It is clear that analytical methods are important to
settle this decade-long debate. One way to depart from the SK
limit is the $1/d$ expansion of Georges {\it et al} \cite{1/d}
supporting the survival of the replica symmetry broken (RSB)
picture of Parisi in high spatial dimensions. Another important question
is whether a spin glass transition exists in a nonzero magnetic field.
Droplet theory gives
a negative answer, while the onset of RSB at the de Almeida-Thouless
(AT) \cite{AT} line is an important feature of Parisi's mean field
theory.
Bray and Roberts \cite{BrRo} have studied the problem using
Wilson's
renormalization group in the replica
symmetric (RS) high-temperature phase,
constraining the fluctuating fields into the
replicon subspace. They did not find any meaningful fixpoint, thus
suggesting that there is no AT-line around 6 dimensions. We feel, however,
that their projected field theory, with all the masses other than
that of the
replicon taken to be infinite, is not sufficient to settle the question
since even the crossover region including the known zero-field fixed-point
\cite{fixpoint}
is impossible to reach in this model with only two cubic coupling constants.
Taking into account all the eight cubic invariants of the permutation group
we would have a large enough parameter space with eight $\varphi^3$
coupling constants and three masses. A renormalisation study, similar
to that of Bray and Roberts, of this model could resolve this problem
(work is in progress in this direction).

In this paper we study the replica field theory corresponding to 
the spin glass phase of the Edwards-Anderson model in zero magnetic
field and just below the transition point. We are mostly interested in
systems with spatial dimensions around the upper critical dimension
$d_u=6$, the case
of higher
dimensions $d>8$
and the problem of the lower critical dimension $d_l$
are left to the end. We have made an extensive perturbation analysis
of the equation of state and found infrared divergent terms at two-loop
order for $d<6$. We will point out that it is the small mass of the bare
propagator, rather than the zero (Goldstone) modes, that
is responsible for
these singularities: it is proportional to $w^2$, the perturbation
parameter. When physical quantities are expanded
in terms of $w^2$, the natural infrared cutoff
of the theory is destroyed. Nevertheless, below six dimensions,
the full theory has
only one mass scale, and we propose a simple
scaling theory where all the infinities coming from the infrared integrals
are absorbed into the amplitudes of scaling forms.

As explained in Ref.~\cite{beyond},
bare coupling constants higher than cubic order must be set to zero
when studying the critical region. The Lagrangean can be split
into a Gaussian and an interaction part:
\[
 {\cal L}={\cal L}^{(2)}+{\cal L}^{(I)} \quad ,
                                               \]
with
\[
   {\cal L}^{(2)}=\frac {1}{2}\sum_{\al<\bet,\gamma<\delta}\sum_{\vec p}
   		  \varphi_{\vec p}^{\al\bet}\left (\tilde {G}^{-1}
		  (p)\right )_{\al\bet,\gamma\delta}
		  \varphi_{-{\vec p}}^{\gamma\delta} \quad ,
		                                    \]
where
\begin{eqnarray}
                   \left (\tilde {G}^{-1}
		  (p)\right )_{\al\bet,\al\bet}&=&p^2-r_0 \quad ,\nonumber\\
		  \left (\tilde {G}^{-1}
		  (p)\right )_{\al\gamma,\bet\gamma}&=&-wQ_{\al\bet}\quad ,
\label{gauss}\end{eqnarray}
and all the other components of $\tilde G^{-1}$ are zero.  $r_0$ and
$w$ are the bare mass and cubic coupling constant, respectively,
while $Q_{\al\bet}$ is the {\em exact} order parameter matrix. Greek
indices above and in the following stand for replica numbers going
from $1$ to $n$, and $n$ is set to zero, according to the replica trick,
at the very end of the calculations. The above Lagrangean is defined
as a functional of the fields $\varphi_{\vec p}^{\al\bet}$ with
zero average values:
\be 
    \langle \varphi_{\vec p}^{\al\bet} \rangle =0\quad .
                                                     \label{zeroave}
						     \ee
(Statistical averages $\langle \ldots\rangle$ are calculated with
the weight $\sim e^{\cal L}$.) Eq.~\ref{zeroave} expresses
the shift of $\varphi^{\al\bet}$, fluctuating now around the exact
$Q_{\al\bet}$, as usual in the description of symmetry broken phases.
$Q_{\al\bet}$ enters also
${\cal L}^{(I)}$:
\[
    {\cal L}^{(I)}={\cal L}^{(1)}+{\cal L}^{(3)}
    						   \]
with						     
\begin{eqnarray*}
           {\cal L}^{(1)}&=&\sqrt{N} \sum_{\al<\bet}
	    \varphi_{\vec {p}=0}^{\al\bet}\left[r_0 Q_{\al\bet}+
	    w(Q^2)_{\al\bet}\right]\quad ,\\
	   {\cal L}^{(3)}&=&\frac{1}{\sqrt N} \, w \sum_{\al<\bet<\gamma}
	    \sum_{\vec p_1,\vec p_2} \varphi_{\vec p_1}^{\al\bet} 
            \varphi_{\vec p_2}^{\bet\gamma}\varphi
            _{-(\vec p_1+\vec p_2)}^{\gamma\al}
	    \quad .
\end{eqnarray*}	    
($N$, the number of sites of the $d$-dimensional hypercubic lattice,
is taken to infinity in the thermodynamic limit.) 

We concentrate on the determination of $Q_{\al\bet}$ near $T_c$
in zero magnetic field where the above Lagrangean is taken with
$r_0=r_0^{(c)}+r$, $r\ll 1$. For such a cubic field theory,
Eq.~\ref{zeroave} can be written as a closed system of two
equations:
\begin{eqnarray}
rQ_{\al\bet}+w(Q^2)_{\al\bet} & = & -\left[ w\int^{\Lambda} \sum_{\rho\ne
\al,\bet} {\hat G}_{\al\rho,\bet\rho}(p)+r_0^{(c)}Q_{\al\bet}\right]
\equiv -Y_{\al\bet}\quad ,\label{eqstate}\\
\langle \varphi_{\vec p}^{\al\bet}\varphi_{-\vec p}^{\gamma\delta}\rangle 
& = &  {\hat G}_{\al\bet,\gamma\delta}(p)\quad ,\label{prop}
\end{eqnarray}
where we used the shorthand notation $\int^{\Lambda}\equiv
\int^{\Lambda}\frac{d^d  p}{(2\pi)^d}$. $\hat G$ is the exact
propagator, while the critical value of $r_0$ is to be determined 
from the implicit equation:
\be 
 r_0^{(c)}= -\lim_{Q\to 0} \frac{1}{Q_{\al\bet}}w\int^{\Lambda}
\sum_{\rho\ne
\al,\bet} {\hat G}^{(c)}_{\al\rho,\bet\rho}(p)\quad, \label{r0c}
\ee
with $\lim_{Q\to 0}$ meaning that all the elements of Q go to zero.
${\hat G}^{(c)}$ depends
on $Q$ and
$r_0^{(c)}$ in a complicated manner. By Eq.~\ref{r0c},
$r_0^{(c)}$ is a
function of the coupling
constant $w$, and it can be calculated perturbatively.

Computing $Q_{\al\bet}$ from Eqs.~\ref{eqstate} and \ref{prop} is a
difficult problem, as it is for any nontrivial field theory.
Without assuming an ansatz for the structure of $Q_{\al\bet}$, it is
completely hopeless to find even an approximate solution. We can try,
however, to follow the traditional way by starting from a mean field-like
equation whose solution is taken as a zeroth order approximation in
a systematic perturbative treatment. The most obvious way of defining
a mean field equation of state from Eq.~\ref{eqstate} is to drop the
loop term $Y_{\al\bet}$. Any solution found must then be
submitted to the
test of stability: the mass operator $\Gamma\equiv {\hat G}^{-1}(\vec p=0)$
should not have negative eigenvalues. The
illusion created by the marginally stable replica symmetric
mean field solution is dispelled
by calculating the
first loop correction (this was made in Ref.~\cite{BrMo}):
Bray and Moore found that the hitherto zero, so called replicon,
eigenvalue moves to a negative value, a clear indication of an effective
quartic coupling (we call it $\tilde u$)
generated by the cubic coupling
at the first loop level \cite{beyond}.

For a meaningful mean field theory, we
have to keep the loop term $Y_{\al\bet}$ in Eq.~\ref{eqstate} and make
a reasonable truncation in Eq.~\ref{prop}. We can avoid instability
by assuming an infinite step RSB structure for $Q_{\al\bet}$.
It turns out that in this case $Q_{\al\bet}$
is highly insensitive to the details of the propagators in the far
infrared region, while the near infrared propagators which determine
$Q_{\al\bet}$ are not influenced by the quartic couplings
(for a discussion of the behaviour of the propagator components in
the far and near infrared, corresponding to the two mass scales, see
Ref.~\cite{beyond}). We have therefore a freedom in truncating
Eq.~\ref{prop}, we can even completely neglect ${\cal L}^{(I)}$ when
computing the average, thus taking $\hat G=\tilde G$. Alternatively,
we may take into account the quartic couplings, $\tilde u$
for instance,
being generated from the interaction
part. In any case, the propagators are the same in the near
infrared, while only the amplitudes are influenced in the far infrared
region giving a relatively moderate divergence ${\hat G}_{\al\rho,\bet\rho}
\sim p^{-3}$. The role of the small mass is to provide an infrared cutoff,
and our mean field theory is certainly meaningful for $d>3$.

Using the tables from Ref.~\cite{beyond} for the near infrared propagators,
we can approximately compute the Parisi order parameter function  in the
vicinity of $T_c$:
\begin{eqnarray}
Q_{\al\bet}=Q(x)&=&\frac{w}{2\tilde u} x \quad, \qquad x=
\al \cap \bet<x_1\quad,
\label{qx}\\
Q(x)=Q_1&=&\frac{1}{2w}r\quad, \qquad x>x_1\quad,\nonumber\\
x_1&=&\frac{\tilde u}{w^2}r \quad, \label{x1}
\end{eqnarray}
where $Q_1$ and $x_1$ stand for the plateau value and breakpoint of
$Q(x)$, respectively. The effective quartic coupling is
\cite{GrMoBr,beyond}:
\be
\tilde u = 12 w^4 \int ^{\Lambda} \frac{1}{p^4(p^2+r)^2} \quad.
\label{u}
\ee
$Q(x)$ in Eq.~\ref{qx} is very much like that of the SK model,
where the coupling constants $w$ and
$\tilde u$ are $1$. We can learn further details about $Q(x)$
by writing a scaling form for $Y_{\al\bet}=Y(x)$. Considering the two mass
scales of the mean field propagators
\begin{eqnarray*}
            m^2_{\rm large}&\sim &r \quad, \\
	    m^2_{\rm small}&\sim & \frac{\tilde u}{w^2}r^2=x_1 r\quad,
\end{eqnarray*}	    
we have in the critical region:
\be
    Y(x)=w \int^{\Lambda} \frac{1}{p^2} f_{\rm mf}(\frac{p^2}{r},
         \frac{p^2}{x_1 r},\frac{x}{x_1})\quad, \label{mfscaling}
\ee
where $f_{\rm mf}$ is a scaling
function. The truncation procedure for the propagator has an effect
on $f_{\rm mf}$ in the momentum range around the small mass
($p^2\sim x_1 r$), but the near infrared region ($p^2\sim r$) and the
power of the far infrared limit ($\sim p^{-1}$) are not influenced
by it.

The mean field approximation presented above is highly nontrivial, and the
order parameter function depends in a very complicated manner on 
the perturbation parameter $w^2$. The origin of this complicated
dependence
can be understood from the fact that the small mass is proportional
to $w^2$:
\be
     m^2_{\rm small}\sim \frac{\tilde u}{w^2}\sim w^2 
     \label{smallmass}
\ee
giving rise to a $w$-dependence of the propagator $\hat G$. At this point
we are tempted to compute $Q(x)$ by perturbation expansion, i.e.\ order
by order in $w^2$. This, however, inevitably leads to infrared divergent
terms since the infrared cutoff provided by the small mass is then
destroyed.
For example, in the expansion of $Q_1$ we have a third order
term like
\[
   wQ_1\sim w^6 r^{d-4} \int^{\Lambda} \frac{1}{p^8}
\]
which is infinite for $d<8$.
Similarly, as we will see later,
when using simple-minded perturbation expansion for the full
theory we may encounter infrared divergent infinite terms
that can, however, build up an infrared cutoff,
thus rendering the theory well-behaved above the lower critical
dimension.

Following standard arguments of the theory of critical phenomena,
we expect a qualitative agreement between full and mean field theory
above the upper critical dimension ($d>6$). Near $T_c$, $Y(x)$
($x=\al\cap\bet$) in Eq.~\ref{eqstate} can be written as:
\be
    Y(x)=w \int^{\Lambda} \frac{1}{p^{2-\eta}} f(\frac{p^2}{r^{2\nu}},
         \frac{p^2}{x_1 r^{2\nu}},\frac{x}{x_1})\quad,
	 \qquad 6<d<8\quad, \label{d>6scaling}
\ee    
where the scaling function $f$ depends now on the coupling $w$ and,
for later reference, the mean field exponents $\eta=0$, $\nu=1/2$
were included. A renormalised $\tilde u$ is now understood in Eq.~\ref{x1}
for the breakpoint of Eq.~\ref{d>6scaling}, preserving its temperature
scaling
\be
      x_1 \sim r^{\frac{d}{2}-3} \quad, \qquad 6<d<8 \quad .
      \label{x1temp}
\ee      

Going below the upper critical temperature seems to be easy now: the
exponents $\eta$ and $\nu$ in Eq.~\ref{d>6scaling} are then nontrivial
and depend on the dimension (they have been computed from the critical
theory up to $O(\epsilon^3)$, $\epsilon=6-d$, in Ref.~\cite{fixpoint}).
The most important point to clarify is the role of $x_1$ in the scaling of
the $Y(x)$ term. Eq.~\ref{x1temp} suggests that $x_1$ becomes temperature
independent in $d=6$. We have made a rather difficult calculation of the
$O(\epsilon)$ correction to the temperature exponent of $x_1$ (see later)
and found that it is zero. The hypothesis of a constant, nonuniversal,
i.e.\ temperature independent and $w$, $\Lambda$ dependent, $x_1$ is
strongly suggested by this finding. The simple scaling picture restored
in $d=6$ would then survive even below the upper critical dimension and
$Y(x)$ would take the form
\be
    Y(x)=w\int^{\Lambda}\frac{1}{p^{2-\eta}} 
    \bar{f}(\frac{p^2}{r^{2\nu}},x)
    \quad, \qquad d<6\quad,
    \label{d<6scaling}
\ee
with the nonuniversal $\bar{f}$
replacing $f$ in Eq.~\ref{d>6scaling}.    

Provided that our assumption of a temperature independent $x_1$ is correct,
a simple form for the order parameter function is strongly
suggested\footnote{This factorization of $r^{\bet}$ from $Q(x)$ was raised
by A.P.~Young some years ago in private communications.}:
\be
    Q(x)=r^{\bet} \bar{Q}(x) \quad, \label{qxscaling}
\ee
where $\bar{Q}(x)$ depends only on
$w$ and $\Lambda$ and for the critical
exponent $\bet$ we have the scaling law
\be
     \bet=\frac{1}{2} \nu (d-2+\eta )\quad . \label{beta}
\ee
We can find full consistency between Eqs.~\ref{d<6scaling} and \ref
{qxscaling} by assuming the limiting form
\be
    \bar{f}(v,x)\sim v^{-\frac{1+\bet}{2\nu}}\quad,\qquad
     v\gg 1  \label{limitingform}
\ee
and using the scaling law Eq.~\ref{beta}. Changing the integration
variable from $\vec p$ to ${\vec p\,}'=\vec p/r^{\nu}$:
\[
   Y(x)=r^{2\bet} w 
        \int^{\Lambda/r^{2\nu}}\frac{1}{p'^{2-\eta}}
	\bar{f}(p'^2,x)
	\cong r^{1+\bet} Y_1(x)+r^{2\bet} Y_2(x) \quad,
\]
where the first term is	the leading contribution coming from the upper
limit
of the integration while the second one is
from the bulk. A $Q(x)$ as
in Eq.~\ref{qxscaling} is then obviously consistent with Eq.~\ref
{eqstate}.

The above theory for
the spin glass phase below $d=6$,
despite its consistency and its clear connection to mean field
theory, does require some analytical support.
We have made a detailed
perturbative computation of the equation of state [\ref{eqstate}] and
our findings can be summarized in the following points:
\begin{enumerate}
\item A Parisi RSB scheme is building up order by order with a
polynomial $Q(x)=ax+cx^3+\ldots$. Consequently,
$\bar{f}(v,x)$ of Eq.~\ref
{d<6scaling} must be also a polynomial in $x$ for any fixed, nonzero
$v$.
\item We computed several logarithms of $r$ to check the above scaling
theory and found
agreement with calculations of the critical exponents in
the critical (massless) phase \cite{fixpoint}. The most important
result comes from the second order (two-loop)
coefficient of $x^3 \log r$
that provides the temperature exponent of the slope $a$ of $Q(x)$
showing that it is equal to $\beta$, at least up to $O(\epsilon)$.
This is consistent with the scaling form in Eq.~\ref{qxscaling}
and
supports the basic
assumption of a temperature-independent $x_1$.

The exponent $\bet$ is originally defined by
\be
     Q_1=C r^{\bet} \label{q1scaling}
\ee
and the one-loop $x\log r$ term gives $\bet$
correctly up to $O(\epsilon)$ ($\bet=1+\frac{\epsilon}{2}$). Furthermore,
using the two-loop $x \log^2 r$
results, exponentiation was checked and
agreement with the known fixed-point value obtained \cite{fixpoint}
(the latter result has been published in \cite{beyond}).

\item We know from exact Goldstone-theorem-like considerations \cite
{goldstone} that an infinite-step RSB solution of the equation of state,
Eqs.~\ref{eqstate} and \ref{prop}, must be marginally stable since the
most dangerous replicon eigenvalues are Goldstone-modes. In contrast to
the RS case \cite{BrMo}, the one-loop calculation
gives indeed a marginally stable spectrum of the mass
operator.

\item The above three points support a Parisi-type
structure of $Q_{\al\bet}$, but now we have to deal with the problem of
the unremovable infrared divergences appearing when a blind
expansion in term of $w^2$ is used. Up to second order, $Y_{\al\bet}$
can be written as
\be
     Y_{\al\bet}=E_{\al\bet}+F_{\al\bet}+O(w^5)\quad, \label{e+f}
\ee
where
\begin{eqnarray}
E_{\al\bet}&\equiv & w\int^{\Lambda} \sum_{\rho\ne
\al,\bet} G_{\al\rho,\bet\rho}(p)+r_0^{(c)}Q_{\al\bet}\quad, \label{e}\\
F_{\al\bet}&\equiv & - w\int^{\Lambda} \sum_{\rho\ne
\al,\bet} \left(G(p)\Gamma^{(1)}(p) G(p)\right)_{\al\rho,\bet\rho}\quad.
\label{f}
\end{eqnarray}
In Eqs.~\ref{e} and \ref{f} we use $G$ instead of $\tilde G$, where
$G$ is defined like $\tilde G$ in Eq.~\ref{gauss}, but with $r$ and
$Q_{\al\bet}^{(0)}$ replacing $r_0$ and $Q_{\al\bet}$, respectively.
$Q_{\al\bet}^{(0)}$
is the leading term of the order parameter:
\[
    Q_{\al\bet}=Q_{\al\bet}^{(0)}+O(w^2)\quad.
\]
The matrix elements of $G^{-1}$ are now all zeroth order, nevertheless
$G$ is not a
free propagator in the usual sense: since $x_1\sim w^2$,
all the sizes of the blocks in the hierarchical construction are also
proportional to $w^2$ and thus $G$ can be expanded in $w^2$:
\[
   G=G^{(0)}+G^{(1)}+\ldots \quad.
\]
As it turns out, $G^{(0)}$ is nothing but the near infrared propagator
whose components were listed in Ref.~\cite{beyond} and further terms
can be calculated by straightforward, although lengthy, perturbation
methods (the component ${G^{(1)}}^{x_1,x}_{1,x_1}$
was needed for the present calculations). 
Two types of infrared divergences with different
origin enter the equation of state:
\begin{itemize} 
\item In Eq.~\ref{f}, where $\Gamma^{(1)}$ is the first order contribution
to the two-point vertex operator $\Gamma \equiv{\hat G}^{-1}$ whose zero
momentum limit is the exact mass operator, the high infrared power of
the two $G$'s could lead to infinite terms. These are, however,
compensated by the zero modes of the mass operator $\Gamma (0)$, as a
direct consequence of exact Ward-identities (very much like in the case
of the Heisenberg model). These Ward-identities follow from the
permutation symmetry of the $n$ replicas which becomes a continuous
symmetry in the infinite step RSB limit \cite{goldstone}. Thus $F_{\al\bet}$
is innocent and well-behaved even below $d=6$.

\item Ward-identities cannot help us to get rid of infrared danger in
$E_{\al\bet}=E(x)$ of Eq.~\ref{e}. To illustrate the problem, we write down
our result for $x=x_1$:
\begin{eqnarray}
     E(x_1) & = & wr^2 \int^{\Lambda} (g_1g_2^2-2g_1^2g_2) +\nonumber\\
            &   & +w^3r^3\frac{4}{5}\left(\int^{\Lambda}g_1^2 g_2^2\right)
	          \int^{\Lambda}(-4g_2^3-5g_1g_2^2-28g_1^2g_2+
	          \nonumber\\
	    &   &+32g_1^3)+O(w^5)\quad, \label{E(x1)}
\end{eqnarray}
where
\[
    g_1\equiv \frac{1}{p^2+r}\quad {\rm and} \quad g_2\equiv
    \frac{1}{p^2} \quad.
\]    
All but one term in Eq.~\ref{E(x1)} give contributions for the computation
of $\bet$ ($\log$'s), checking of exponentiation ($\log^2$'s) and even to
the amplitude $C$ in Eq.~\ref{q1scaling} (simple powers) up to $O(\epsilon
^2)$. The term $-4g_2^3$ is, however, infrared divergent, leading to an
infinite contribution to $C$ at order $\epsilon^2$.
\end{itemize}
\end{enumerate}

The origin of the infrared terms is identical to that
in the mean
field case: $ \sum_{\rho\ne
\al,\bet} G_{\al\rho,\bet\rho}(p)$ is $p^{-3}$ in the zero momentum limit, but
when expanding it in $w^2$, high infrared powers are generated order by
order. These enter the amplitudes of scaling forms like Eqs.~\ref
{d<6scaling}, \ref{limitingform} and \ref{q1scaling}, suggesting that
amplitudes cannot be expanded without producing infinite terms.

The above scaling theory of the equation of state can be extended to the
propagator $\hat G$ (a first attempt in this
direction has been made in \cite{low,Domenico}).
Indeed, it is plausible (and calculations on the
Gaussian level suggest) that {\em any} component of $\hat G$ has the
scaling form $p^{-(2-\eta)} g(p^2/r^{2\nu},x,\ldots)$ where the scaling
function $g$ is, of course, different for different components and $\ldots$
stands for other possible overlaps of the given propagator component. We
record here the explicit formula for the diagonal element
$\hat G^{x,x}_{1,1}$ which is the Fourier transform of the overlap of the
spin-spin correlation function (for the physical meaning of the propagators,
see Ref.~\cite{beyond}):
\be
    G^{x,x}_{1,1}(p)=\frac{1}{p^{2-\eta}} g_{\rm diag}(\frac{p^2}{r^{2\nu}},
    x)\quad, \qquad d<6\quad.  \label{diagonal}
\ee
For $x=0$, it is also the most accessible to
numerical simulations \cite{num}.

The theory presented above is based on the assumption of an
ultrametrically organized Parisi RSB ansatz.
Its analytical justification requires
perturbation methods like the $\epsilon$-expansion
to compute critical exponents ($\eta$, $\nu$ and $\bet$) and check
exponentiation. Amplitudes, however, are infected by infrared divergences,
and the resummation of these infinities seems to be a hard task. The
difficulties in using a perturbative approach for a system with a complex
phase space structure has been discussed in the
context of the random manifold
problem in Ref.~\cite{manifold}. Numerical methods would be valuable
to check formulae like Eqs.~\ref{qxscaling} and \ref{diagonal}, and compare
the measured exponents with those obtained at or above $T_c$. For this
purpose, we make further assumptions for $G^{x,x}_{1,1}$ in Eq.~\ref
{diagonal} when $x=0$, namely
\begin{eqnarray}
                 g_{\rm diag}(v,0) & \to & {\rm constant}\quad,
		 \qquad v\to \infty\quad, \nonumber\\
		 g_{\rm diag}(v,0) & \sim & v^{-\frac{\lambda}{2}}
		 \quad, \qquad v\to 0\quad. \label{lambda}
\end{eqnarray}
The value of the exponent $\lambda$ has an important role in the 
determination of the lower critical dimension $d_l$ (see below).

Up to now, we confined our discussion to systems with $d<8$. Above eight
dimensions we must keep the bare quartic coupling $u$ to avoid instability
of the RS solution, and all the formulae containing $\tilde u$ are still
valid with $u$ replacing $\tilde u$. $Y(x)$ in Eq.~\ref{eqstate} has the
mean field form like in Eq.~\ref{d>6scaling} with $\eta =0$, $\nu =1/2$
and $x_1=ur/w^2 \sim r$. The scaling function $f$ depends now on $w$,
$u$, $\Lambda$ and, in high enough dimension, possibly on other higher
order bare couplings, and on the l.h.s of Eq.~\ref{eqstate}
terms like
$\frac{2}{3}u Q_{\al\bet}^3$ appear.
The theory is basically nonperturbative in the
sense that a perturbation expansion in the bare couplings destroys the
infrared cutoff at $p\sim r$.

The problem of the lower critical dimension $d_l$ where the above scaling
theory blows up, can be approached from two different directions:
\begin{itemize}
\item
In the case of any critical transition,
$T_c$ can reach zero temperature in low
enough dimension which, in a field
theoretic representation, manifests itself in a
divergent $r_0^{(c)}$. From the observation that the second term
$Q_{\al\bet} r_0^{(c)}$ in $Y_{\al\bet}$ (see Eq.~\ref{eqstate}) is to cancel
the leading contribution of the first one when $p^2\gg r$, we can put
Eq.~\ref{r0c} into the simple form consistent with our scaling picture:
\be
    r_0^{(c)}\sim \lim_{r\to 0} r^{-\bet}\int^{\Lambda}
    \frac{1}{p^{2-\eta}}\tilde{f}
    \left(\frac{p^2}{r^{2\nu}}\right)
    \sim \int^{\Lambda}
    \frac{1}{p^{2-\eta+\bet/\nu}}\quad, \label{r0cscaling}
\ee
where the asymptotic form $\tilde{f}(v)
\sim v^{-\bet/2\nu}$, $v\to \infty$, has been
used. This can now be compared with the subleading term of Eq.~\ref
{limitingform} remaining after the cancellation mentioned above. Using
the scaling law of Eq.~\ref{beta}, we arrive at the conclusion that the
integral of Eq.~\ref{r0cscaling} is convergent provided $d>2-\eta$. The
lower critical dimension following from this argument satisfies the
equation
\be 
    d_l^{(1)}=2-\eta(d_l^{(1)})\quad. \label{dl1}
\ee
\item
In systems with Goldstone modes, the high infrared powers
of the propagators lead to divergences in local quantities at some
$d_l^{(2)}$, thereby signalling the breakdown of the ordered phase.
When
$d_l^{(2)}>d_l^{(1)}$, it is the lower critical dimension (a situation
known from the XY and Heisenberg models where $d_l=2$). As it was suggested in
Ref.~\cite{low}, we must consider the
most dangerous component $G^{0,0}_{1,1}$
whose far infrared behaviour can be obtained from
Eqs.~\ref{diagonal} and \ref{lambda}:
\be
     G^{0,0}_{1,1}(p) \sim p^{-(2-\eta+\lambda)}\quad,
     \qquad p\to 0\quad. \label{farinfrared}
\ee     
From this we have
\be
     d_l^{(2)}=2-\eta(d_l^{(2)})+\lambda(d_l^{(2)})\quad.
     \label{dl2}
\ee
In Ref.~\cite{low}, after an incomplete perturbative analysis, it
was argued that $\lambda=\bet /\nu$, giving rise to $d_l^{(2)}=
d_l^{(1)}$, leading to a $d_l$ which is definitely smaller than three.
Numerical estimate of the exponent $\lambda$ for $d=3$ can be found in
Ref.~\cite{num}, namely $\lambda=2\bet/\nu -\al$ with $\al \approx 0.5$ which
gives $\lambda\approx 0.1$.
This result should be taken with some care given the small size of the
system considered. If accepted, then it gives
a smaller $\lambda$ than the previous one:
$\lambda=\bet/\nu\approx 0.3$, $d=3$, and is thus even more consistent with
$d_l<3$. ($d_l=2.5$ was suggested by Franz {\it et al} \cite{Franz}
as the lower critical dimension where replica symmetry is restored.) 
Nevertheless, we cannot exclude the possibility $d_l=3$, and more
numerical and analytical work \cite{prep}
would be useful to settle this problem.
\end{itemize}

\section*{Acknowledgements}
This work has been supported by the Hungarian Science
Fund (OTKA), Grant Nos.\ T017493 and T019422,
and by the French Ministry of Foreign Affairs
(CEA/MAE \# 43).

\end{document}